\documentclass[runningheads]{llncs}
\usepackage{hyperref}
\usepackage{aodv}
\usepackage{url}
\usepackage{amsfonts,amsmath,amssymb}
\usepackage{algorithm,algorithmic}
\floatname{algorithm}{Table}
\renewcommand{\listalgorithmname}{List of Processes}
\usepackage{graphicx}
\usepackage{color}
\usepackage{pdfsync}
\usepackage[caption=false]{subfig}
\usepackage{ifthen}
\usepackage{xspace}
\usepackage{rotating}
\usepackage{multirow}

{\bfseries}{\itshape}%
\newcommand{\numberedthing}[2]{%
     \newtheorem{#1}[common]{#2}{\bfseries}{\itshape}}%
\numberedthing{prop}{Proposition}%

\newcommand{\awn}{AWN\xspace}
\newcommand{\uppaal}{UPPAAL\xspace}

\begin{document}%
\title{Automated Analysis of AODV using \uppaal
\texorpdfstring{\thanks{First steps towards this analysis appeared in~\cite{FGHMPT11}.}}{}}%

\authorrunning{Fehnker, van Glabbeek, H\"ofner, McIver, Portmann \& Tan}
  \author{
  Ansgar Fehnker\inst{1,4},
  Rob van Glabbeek \inst{1,4},
  Peter H\"ofner\inst{1,4},
  Annabelle McIver\inst{2,1},
  Marius Portmann\inst{1,3}\and
  Wee Lum Tan\inst{1,3}
  }
\institute{
  NICTA\and
  Department of Computing, Macquarie University\and
  School of ITEE, The University of Queensland\and
  Computer Science and Engineering, University of New South Wales
}

\maketitle 
\setcounter{footnote}{0}
\begin{abstract}
This paper describes an automated, formal and rigorous analysis of the
Ad hoc On-Demand Distance Vector (AODV) routing protocol, a popular
protocol used in wireless mesh networks.

\hspace{1em}
We give a brief overview of a model of AODV implemented in the \uppaal model checker.
It is derived from a process-algebraic model which reflects precisely the intention of AODV and 
accurately captures the protocol specification.
Furthermore, we describe experiments carried out to explore AODV's behaviour in all
network topologies up to $5$~nodes. We were able to automatically locate
problematic and undesirable behaviours. This is in particular useful {to discover} protocol
limitations and {to develop} improved variants.
This use of model checking as a diagnostic tool complements other 
formal-methods-based protocol modelling and verification techniques, such as process algebra.

\end{abstract}
\hfuzz2pt 

\section{Introduction}\label{sec:intro}
Route finding and maintenance are critical for the performance of
networked systems, particularly when mobility can lead to highly
dynamic and unpredictable environments; such operating contexts are
typical in wireless mesh networks (WMNs).  Hence correctness and good
performance are strong requirements of routing algorithms. The Ad hoc
On-Demand Distance Vector (AODV) routing protocol~\cite{rfc3561} is a
widely used routing protocol designed for WMNs and mobile ad hoc
networks (MANETs).  It is one of the four protocols defined in an RFC
(Request for Comments) document by the IETF MANET working group.  AODV
also forms the basis of new WMN routing protocols, like the
upcoming IEEE 802.11s wireless mesh network
standard~\cite{IEEE80211s}.

Usually, routing protocols are optimised to achieve key objectives
such as providing self-organising capability, overall reliability and
performance in typical network scenarios. Additionally, it is
important to guarantee protocol properties such as loop freedom for
{\em all\/} scenarios, including non-typical, unanticipated ones.
This is particularly relevant for highly dynamic MANETs and WMNs.

The traditional approaches for the analysis of MANET and WMN routing
protocols are simulation and test-bed experiments. 
While these are important and valid methods for protocol
evaluation, there are limitations: they are resource intensive and
time-consuming.  The challenges of extensive experimental evaluation
are illustrated by recent discoveries of limitations of protocols that
have been under intense scrutiny over many years. An example is~\cite{MK10}.

We believe that formal methods in general and model checking in
particular can help in this regard.  Model checking is a powerful
method that can be used to validate key correctness properties in
finite representations of a formal system model. In the case that a
property is found not to hold, the model checker produces evidence for
the fault in the form of a ``counter-example" summarising the
circumstances leading to it. Such diagnostic information provides
important insights into the cause and correction of these failures.

In \cite{TR11}, we specified the AODV routing protocol in the process
algebra \awn.  The specification follows well-known programming constructs and lends
itself well for comparison with the original specification of the
protocol in English.  Based on such a comparison we believe that the
\awn model provides a complete and accurate formal specification of
the core functionality of AODV\@.  In developing the formal
specification, we discovered a number of ambiguities in the IETF RFC
\cite{rfc3561}.  Our process algebraic formalisation captures these by
several interpretations, each with slightly different {\awn} code.

In this paper we follow an interpretation of the RFC, which we believe to be the closest
to the spirit of the  AODV routing protocol. We show how to obtain
executable versions of this \awn specification, in the language of
the \uppaal model checker~\cite{uppaal04,LPY97}. By deriving the
UPPAAL model from the \awn model, the accuracy of the \awn model is
transferred to the \uppaal model.  

The executable \uppaal model is used to confirm and discover the
presence of undesirable behaviour. We check important
properties against all topologies of up to $5$ nodes, which also
includes dynamic topologies with one link going up or down. This
exhaustive search confirmed known and revealed new problems of AODV, and let us
quantify in how many topologies a particular error can
occur. Subsequently, the same experiments for
modifications of AODV showed the proposed modifications can all
but eliminate certain problems for static topologies, and
significantly reduce them for dynamic topologies. The automated
analysis of routing protocols presented in this paper combined with
formal reasoning in \awn provides a powerful tool for the development
and rigorous evaluation of new protocols and variations, and
improvements of existing ones.

\section{Ad hoc On-Demand Distance Vector Routing Protocol}\label{sec:AodvOverview}
\subsection{The Basic routine}
\begin{figure}[t]
 \begin{center}
   \begin{tabular}[b]{r@{}l@{\hspace{13mm}}r@{}l@{\hspace{13mm}}r@{}l}
   (a)&
   \includegraphics[scale=1]{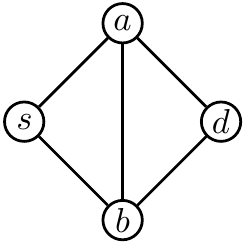}&
   (b)&
   \includegraphics[scale=1]{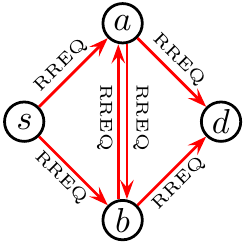}&
   (c)&
   \includegraphics[scale=1]{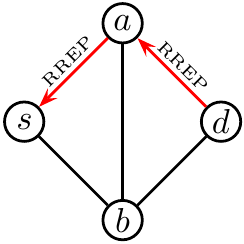}
   \end{tabular}
   \caption{Example network topology}
   \label{fig:topology}
 \end{center}
 \vspace{-3em}
\end{figure}

AODV~\cite{rfc3561} is a widely used routing protocol designed for WMNs and {MANETs}. It is a reactive routing protocol, where the route between a source and a destination node is established on an on-demand basis. A route discovery process is initiated when a source node $s$ has data to send to a destination node~$d$, but has no valid corresponding routing table entry. In this case, node $s$ broadcasts a route request (RREQ) message in the network. The RREQ message is re-broadcast and forwarded by other intermediate nodes in the network, until it reaches the destination node $d$ (or an intermediate node that has a valid route to node $d$). Every node that receives the RREQ message will create a routing table entry to establish a \emph{reverse route} back to node $s$. In response to the RREQ message, the destination node~$d$ (or an intermediate node that has a valid route to node~$d$) unicasts a route reply (RREP) message back along the previously established reverse route. At the end of this route discovery process, an end-to-end route between the source node~$s$ and destination node~$d$ is established. Usually, all nodes on this route have a routing table entry to both the source node~$s$ and destination node $d$.
An example topology, {indicating which nodes are in transmission range
of each other,} as well as  the flow of RREQ and RREP messages, is
given in \Fig{topology}.
In the event of link and route breaks, AODV uses route error (RERR)
messages to inform affected nodes. Sequence numbers are another
important aspect of AODV, and are used to indicate the freshness of
routing table entries for the purpose of preventing routing loops.

\subsection{Process Algebraic Model of AODV}\label{ssec:processalgebra}
The process algebra \awn~\cite{ESOP12,TR11}
has been developed specifically for modelling WMN routing protocols. It is designed in a way to be easily readable and
treats three necessary features of WMNs protocols:
{\em data structures}, {\em local broadcast}, and  {\em conditional unicast}.
Data structures are used to model routing tables etc.;
local broadcast models message sending to {\em all} directly
connected nodes; and 
conditional unicast models the
message sending
to one particular node and chooses a continuation process dependent on whether the
message is successfully delivered.

In AWN, delivery of broadcast messages is ``guaranteed'', i.e., they
are received by any neighbour that is directly connected.  The
abstraction to a guaranteed broadcast enables us to interpret a
failure of message delivery (under assumptions on the network
topology) as an imperfection in the protocol, rather than as a consequence of unreliable communication. Section~\ref{subsec:non-optimal-route}, for example, describes a simple
network topology and a scenario for which AODV fails to discover a
route, even if broadcast is
guaranteed. The failure is a shortcoming of the
protocol itself, and cannot be excused by unreliable communication.

Conditional unicast models an abstraction of an
ac\-know\-ledg\-ment-of-receipt mechanism that is typical for unicast
communication but absent in broadcast communication, as implemented by
the link layer of relevant wireless standards such as IEEE 802.11. The
{\awn} model captures the bifurcation depending on the
success of the unicast, while abstracting from all implementation
{details}.
\pagebreak[3]

In \cite{TR11}, we used \awn to model AODV according to the IETF
RFC~\cite{rfc3561}.  The model captures all core functionalities as
well as the interface to higher protocol layers via the injection and
delivery of application layer data, and the forwarding of data packets
at intermediate nodes.  Although the latter is not part of the AODV
protocol specification, it is necessary for a practical model of any
reactive routing protocol where protocol activity is triggered via the
sending and forwarding of data packets. In addition, our model
contains neither ambiguities nor contradictions, both of which are often present 
in specifications written in natural languages, such as in the RFC3561
(see e.g.~\cite{TR11}).

The \awn model of AODV contains a main process, called {\AODV}, for every
node of the network, which handles messages received and calls the
appropriate process to handle them. The process also handles the forwarding of any queued data
packet if a valid route to its destination is known. Four other
processes handle one particular message type each, like {RREQ}. The
network as a whole is modelled as a parallel composition of these
processes.  Special primitives allow us to express whether two nodes
are connected.  Full details of the process algebra description on
which our \uppaal model is based can be found in \cite{TR11}.

\section{Modelling AODV in \uppaal}\label{sec:uppaal}
\uppaal~\cite{uppaal04,LPY97} is an established model checker for \emph{networks of timed automata}, used in particular for protocol verification.
We use \uppaal for the following reasons: 
(1) \uppaal provides two synchronisation mechanisms---binary and broadcast synchronisation, which translate to uni- and broadcast communication;
(2) it provides common data structures, such as arrays and structs, and a C-like programming language to define updates on these data structures;
(3) in the future, \awn (and therefore also our models) will be extended with time and probability---\uppaal provides mechanisms and tools for both.

Our process-algebraic model of AODV has been used to prove essential
properties, such as loop freedom for popular interpretations of \cite{rfc3561}---independent of a particular topology.  The \uppaal model is
derived from the \awn specification that comes closest to the spirit of the  AODV routing protocol. 

\SSect{translation} explains the translation and the simplifying assumptions in detail.

\subsection{\uppaal Automata}

Since our models do not yet use time (or probabilities) they are simply \emph{networks of automata} with guards. The state of the system  is determined, in part, by the values of data variables that can be either shared between automata, or local. We assume a data structure with several types, variables ranging over these types, operators and  predicates. Common Boolean and arithmetic expressions are used to denote data values and statements about them.

Each automaton is a graph, with locations, and edges between
locations. Every edge has a guard, optionally a synchronisation label,
and an update. Synchronisation occurs via so-called channels; for each
channel $a$ there is one label $a!$ to denote the sender, and $a?$ to
denote the receiver. Transitions without labels are internal; all other transitions use one of two types of synchronisation.

In \emph{binary handshake} synchronisation, one automaton having an edge with a label that has the suffix $!$ synchronises with another automaton with an edge having the same label that has a $?$-suffix. These two transitions synchronise when both guards are true in the current state, and only then. When the transition is taken both locations change, and the updates will be applied to the state variables; first the updates on the $!$-edge, then the updates on the $?$-edge. If there is more than one possible pair, then the transition is selected non-deterministically.

In \emph{broadcast} synchronisation, one automaton with a $!$-labelled edge synchronises
with a set of other automata that all have an edge with a matching $?$-label. The
initiating automaton can change its location, and apply its update, if the guard on its edge
evaluates to true. It does not require a second synchronising automaton. Automata with a
matching $?$-labelled edge have to synchronise if their guard is currently true. They change their location and update the state. The automaton with the $!$-edge will update the state first, followed by the other automata in some lexicographic order. If more than one automaton can initiate a transition on an $!$-edge, the choice will be made non-deterministically.

\subsection{From \awn to \uppaal}\label{ssec:translation}
Every node in the network is modelled as a single automaton, each having its own data structures such as a routing table and message buffer. 
The implementation of the data structure defined in \awn is straightforward, since both \awn and \uppaal allow C-style data structures.  
A routing table \texttt{rt} for example is an array of entries, one entry for every node. An entry is given by the data type
\vspace{-1pt}

{\footnotesize
\begin{verbatim}
     typedef struct                    
     { SQN dsn;     //destination sequence number
       bool flag;   //validity of a routing table entry
       int hops;    //distance (hop count) to the destination
       IP nhop;     //next hop (is 0 if no route)
     } rtentry;
\end{verbatim}}
\noindent where \verb+SQN+ denotes a data type for sequence numbers and \verb+IP+ denotes one for all IP address.  In our model, these types are mapped to integers.

The local message buffer is modelled as an array
\texttt{msglocal}. \uppaal will warn if during model checking an
out-of-bounds error occurs, i.e., if the array was too small. Each
message is a struct with fields \texttt{msgtype} which can take values
{\PKT}, {\RREQ}, {\RREP}, or {\RERR}, integer \texttt{hops} for the
distance from the originator of the message, sequence number \texttt{rreqid} to identify a route request, a destination IP \texttt{dip}, a destination sequence number \texttt{dsn}, an originator IP \texttt{oip}, an originator sequence number \texttt{osn}, and a sender IP \texttt{sip}. The model contains functions \texttt{addmsg}, \texttt{deletemsg} and \texttt{nextmsg}, to add a message, delete a message, or to return the type of the next message in the buffer.
\pagebreak[3]

Connections between nodes are determined by a \emph{connectivity graph}, which is
specified by a Boolean-valued function \texttt{isconnected}. This graph presents one
particular topology and is not derived from our \awn specification, since the
specification is valid for {\em all\/} topologies. Communication  is modelled as an atomic synchronised transition between a sender, on an $!$-edge, with a receiver, on a matching $?$-edge. The guard of the sender depends on local data, e.g.~buffer and routing table, while the guard of the receiver is \texttt{isconnected}.
This means that in broadcast communication the sender will take the transition regardless of \texttt{isconnected}, while disconnected nodes will not synchronise. In unicast communication the transition is blocked if the intended recipient is not connected, but there is a matching broadcast transition that sends an error message in this case. When the transition is taken, the sender copies its message to a global variable \texttt{\small msgglobal}, and the receiver copies it subsequently to its local buffer \texttt{msglocal}.

\begin{table}[t]\vspace*{-2em}
  \algsetup{linenodelimiter=.,linenosize=\tiny}
  \begin{algorithm}[H]
    {
      \scriptsize
      \caption{Excerpt of \awn spec for AODV. A few cases for RREQ handling.}
      \label{pro:rreq}
      \begin{algorithmic}[1]
        \DEFPROCESS{\AODV}{\ip,\sn,\rt,\rreqs,\queues}
		\COMLINE{depending on the message on top of the message queue, the node calls different processes}	
		\STATE{$\dots$}
		\IF{$\msg = \rreq{\hops}{\rreqid}{\dip}{\dsn}{\oip}{\osn}{\sip}
			\wedge (\oip,\rreqid)\in\rreqs$}			
			\COMLINE{silently ignore RREQ, i.e. do nothing, except update the entry for the sender}
			\UPD{\rt:=\upd{\rt}{(\sip,0,\val,1,\sip)}}\ .           \COMMENT{update the route to \sip}
 \aodvL{\ip}{\sn}{\rt}{\rreqs}{\queues} 
		\ELSIF{$\msg = \rreq{\hops}{\rreqid}{\dip}{\dsn}{\oip}{\osn}{\sip}
			\wedge (\oip,\rreqid)\not\in\rreqs) 
			\wedge \dip=\ip$}
                        \COMLINE{answer the RREQ with a RREP}
			\UPD{\rt:=\upd{\rt}{(\oip,\osn,\val,\hops+1,\sip)}}	\COMMENT{update the routing table} 
			\label{rreq:line6A}
			\UPD{\rreqs:=\rreqs\cup\{(\oip,\rreqid)\}}		\COMMENT{update  the array of already seen RREQ}
																\label{rreq:line8}
			\UPD{\sn:=\max(\sn,\dsn)}	                        \COMMENT{update the sqn of \ip}	
			\UPD{\rt:=\upd{\rt}{(\sip,0,\val,1,\sip)}}              \COMMENT{update the route to \sip}	
			\STATE\unicast{\nhop{\rt}{\oip}}{{\rrep{$0$}{\dip}{\sn}{\oip}{\ip}}}\ . 											\label{rreq:line14a}								
					\aodvL{\ip}{\sn}{\rt}{\rreqs}{\queues}
		\ELSIF{$\msg = \rreq{\hops}{\rreqid}{\dip}{\dsn}{\oip}{\osn}{\sip}
			\mathop\wedge (\oip,\rreqid)\not\in\rreqs) 
			\mathop\wedge \dip\neq \ip 
                            \mathop\wedge
                            \newline\hspace*{1.8em}
                             (\dip\mathbin{\not\in}\akD{\rt} \vee \sqn{\rt}{\!\dip} <  \dsn \vee\sqnf{\rt}{\!\dip}\mathbin=\unkno)$}		
                        \COMLINE{forward RREQ}
			\UPD{\rt:=\upd{\rt}{(\oip,\osn,\val,\hops+1,\sip)}}	\COMMENT{update routing table} 
			\label{rreq:line6B}
			\UPD{\rreqs:=\rreqs\cup\{(\oip,\rreqid)\}}		\COMMENT{update the array of already seen RREQ} 
			\UPD{\rt:=\upd{\rt}{(\sip,0,\val,1,\sip)}}              \COMMENT{update the route to the sender}	
			\broadcast{\rreq{$\hops+1$}{\rreqid}{\dip}{$\max(\sqn{\rt}{\dip},\dsn)$}{\oip}{\osn}{\ip}}\ .					
			\aodvL{\ip}{\sn}{\rt}{\rreqs}{\queues}
		\ELSIF{$\rreq{\hops}{\rreqid}{\dip}{\dsn}{\oip}{\osn}{\sip}
			\wedge \dots$}
			\STATE{$\dots$}		
		\ENDIFii

	\end{algorithmic}
    }
  \end{algorithm}

\vspace*{-5em}
\end{table}

AODV uses unicast for {RREP} and {PKT} messages, and broadcast for {RERR} and {RREQ} messages. To model unicast, the \uppaal model has one binary handshake channel for every pair of nodes. For example, \texttt{rrep[i][j]} is used for transitions modelling the sending of a route reply from node $i$ to $j$. To model broadcast, we use one broadcast channel for every node. For example, \texttt{rreq[i]} is used for the route requests of node $i$. To model new packets from~$i$ to $j$, generated by the user layer, the model contains a channel \texttt{newpkt[i][j]}.

The \awn model of Table \ref{pro:rreq} is an excerpt of the AODV
specification presented in~\cite{TR11}---the full specification and a detailed explanation
can be found there.
The excerpt presented here differs slightly from the original model:\footnote{It can be shown that the model presented here behaves identical to the AWN model in [4]; in other words, they are behavioural equivalent.}
(1) we abstract from \emph{precursors}, an additional data structure that is maintained by AODV
(2) the model in \cite{TR11} uses $6$ different processes; here processes are inlined into the body of the main AODV process. This 
reduces the number of process{es} to one and yields an automaton with one control location;
(3) the model in \cite{TR11} uses nesting of conditions and updates, while this model has been flattened to correspond more closely with the limitations of the \uppaal 
syntax---in \uppaal the \emph{guards} are evaluated before any update, \awn has no such restriction.

\begin{table}[t]\vspace*{-2em}
  \algsetup{linenodelimiter=.,linenosize=\tiny}
  \begin{algorithm}[H]
    {
      \scriptsize
      \caption{Excerpt of \uppaal model. A few cases for RREQ handling.}
      \label{pro:xta}
      \begin{algorithmic}[1]
        \STATE $\ldots$ 
\STATE \verb|aodv -> aodv {|
\STATE \textbf{guard}\verb| nextmsg()==RREQ && rreqs[msglocal[0].oip][msglocal[0].rreqid]; |
\STATE \textbf{sync}\verb|  tau[ip]?;|
\STATE \textbf{assign}\verb| sipupdate(), deletemsg();  },|
\STATE \verb|aodv -> aodv {|
\STATE \textbf{guard}\verb| nextmsg()==RREQ&&!rreqs[msglocal[0].oip][msglocal[0].rreqid]&&msglocal[0].dip==ip;|
\STATE \textbf{sync}\verb|  rrep[ip][oipnhop()]!;|
\STATE \textbf{assign}\verb| updatert(msglocal[0].oip,msglocal[0].osn,1,msglocal[0].hops+1,msglocal[0].sip),|
\STATE \verb|       rreqs[msglocal[0].oip][msglocal[0].rreqid]=1,|
\STATE \verb|       sn=max(sn,msglocal[0].dsn),|
\STATE \verb|       sipupdate(),|
\STATE \verb|       msgglobal=createrep(0,msglocal[0].dip,sn,msglocal[0].oip,ip), deletemsg();  },|
\STATE \verb|aodv -> aodv { |
\STATE \textbf{guard}\verb| nextmsg()==RREQ&&!rreqs[msglocal[0].oip][msglocal[0].rreqid]&&msglocal[0].dip!=ip| 
       \verb+       && (!rt[msglocal[0].dip].flag || msglocal[0].dsn>rt[msglocal[0].dip].dsn+ 
       \verb+       || rt[msglocal[0].dip].dsn==0);+
\STATE \textbf{sync}\verb|  rreq[ip]!;|
\STATE \textbf{assign}\verb| updatert(msglocal[0].oip,msglocal[0].osn,1,msglocal[0].hops+1,msglocal[0].sip),|
\STATE \verb|       rreqs[msglocal[0].oip][msglocal[0].rreqid]=1,|
\STATE \verb|       sipupdate(),|
\STATE \verb|       msgglobal=createreq(msglocal[0].hops+1,msglocal[0].rreqid,msglocal[0].dip,|
       \verb|       max(msglocal[0].dsn, rt[msglocal[0].dip].dsn),msglocal[0].oip,msglocal[0].osn,ip),|
\STATE \verb|       deletemsg();  },|
\STATE $\ldots$

	\end{algorithmic}
    }
  \end{algorithm}

\vspace*{-5em}
\end{table}

Table \ref{pro:rreq} depicts three of the cases in the AWN model for handling route requests. In each, a condition is checked,  the routing tables and local data are updated, and it returns to the main AODV process $\aodv{\ip}{\sn}{\rt}{\rreqs}{\queues}$. Table \ref{pro:xta} shows the corresponding edges from the \uppaal model, one edge for every case. Like the \awn model, which goes from the process $\AODV$ to $\AODV$, the \uppaal model will go from control location \texttt{aodv} to itself (Lines~2, 6 and~14).

Each edge evaluates a guard  in Lines 3, 7 and 15 in Table \ref{pro:xta}. These line numbers, and the line numbers mentioned in the remainder of this section correspond to the same line number in Table \ref{pro:rreq}. Whenever the \awn specification uses set membership ($(\oip,\rreqid)\mathop\in\rreqs$), the \uppaal model uses a 2-dimensional Boolean array \texttt{rreqs} to encode membership;
whenever the \awn model uses a flag to denote a \emph{known}
sequence number ($\sqnf{\rt}{\!\dip}\mathop=\unkno$), the \uppaal model compares with a distinguished value (\texttt{rt[msglocal[0].dip].dsn==0}).

Depending on whether a case requires no transmission, unicast, or
broadcast, the \uppaal model synchronises on a \texttt{tau}, a binary,
or a broadcast channel (Lines 4, 8 and 16). The \texttt{tau} channel
for internal transitions allows for optimisations; it could have been
left empty. We discuss this later in this section.

After synchronisation the state is updated. For all route request messages we update the routing table for the sender $\sip$ (Lines 5, 12 and 19). The fact that the message was received means that sender $\sip$ is one hop away. Except for the first case (Lines 4) the routing table is updated (Lines 9 and 17), and the route request is added to the set of processed route requests (Lines 10 and 18). In case that a node receives a request, and it is the destination, it increments its sequence number, if necessary (Line~11), before it sends a route reply.

The last two steps in the \uppaal model that complete a transmission first create a new message and copy it to the global variable \texttt{msgglobal} (Lines~13 and 20), and then delete the first element of the local message buffer. In the \awn model, these steps are part of the communication primitives.
 
The full \uppaal models a node by an automaton with one control location and $26$ edges:
$19$ cases for processing the different routing messages,
four cases for receiving routing messages---one case for each type---two cases for sending data packets, and 
one case for handling new data packets. 
The case distinction is complete, i.e at least one transition is enabled and process messages if the buffers and queues are not empty. 

Both the \uppaal and the \awn model maintain a FIFO buffer for incoming messages. Any newly generated message only depends on the content of messages  previously received. This implies that the timing of internal transitions that discard incoming messages is not relevant for route discovery. The \uppaal model exploits this fact and assigns a higher priority to internal transitions. To implement priorities we labelled those transitions \texttt{tau}. This is is an effective measure to reduce the state space, at the expense that \uppaal 
is now unable to check liveness properties; for this paper this is not a limitation, as all properties can be expressed as safety properties.

\section{Experiments}\label{sec:experiments}
Our automated analysis of AODV considers $3$ properties that relate to ``route discovery'' for all topologies up to $5$ nodes, with up to one topology change, and scenarios with two new data packets.

\subsection{Scenarios and Topologies}
The experiments consider scenarios with two initial data packets in networks with up to $5$ nodes. 
Initially all routing tables and buffers are empty. The originator and the destination of the data packets are identified as nodes $A$, $B$, or $C$.
\begin{figure}[t]
 \centering \fbox{\hspace*{-1.5mm} 
\includegraphics[width=0.23\linewidth]{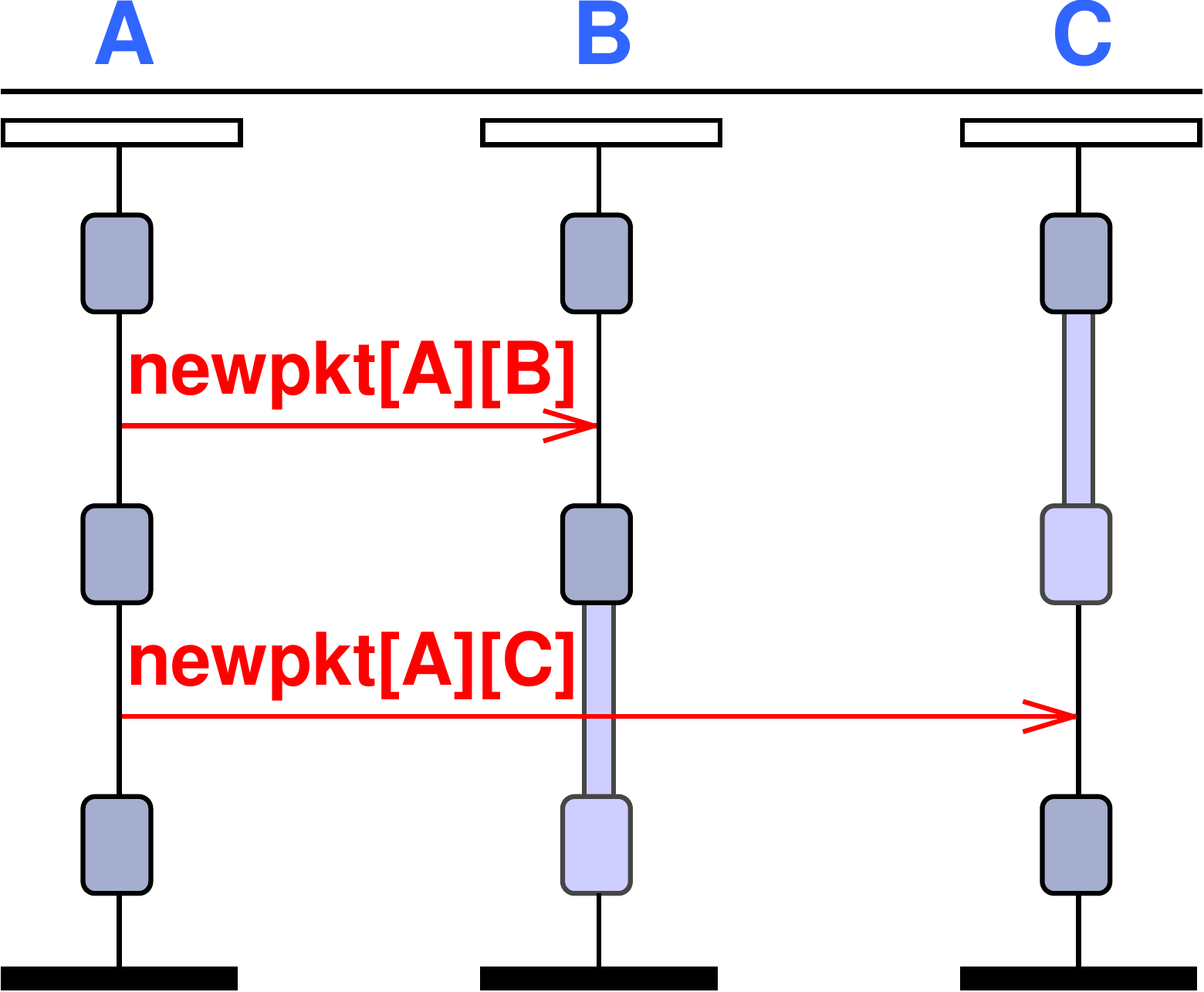}\hspace*{-1.5mm} 
}
\hfill\fbox{\hspace*{-1.5mm} 
\includegraphics[width=0.23\linewidth]{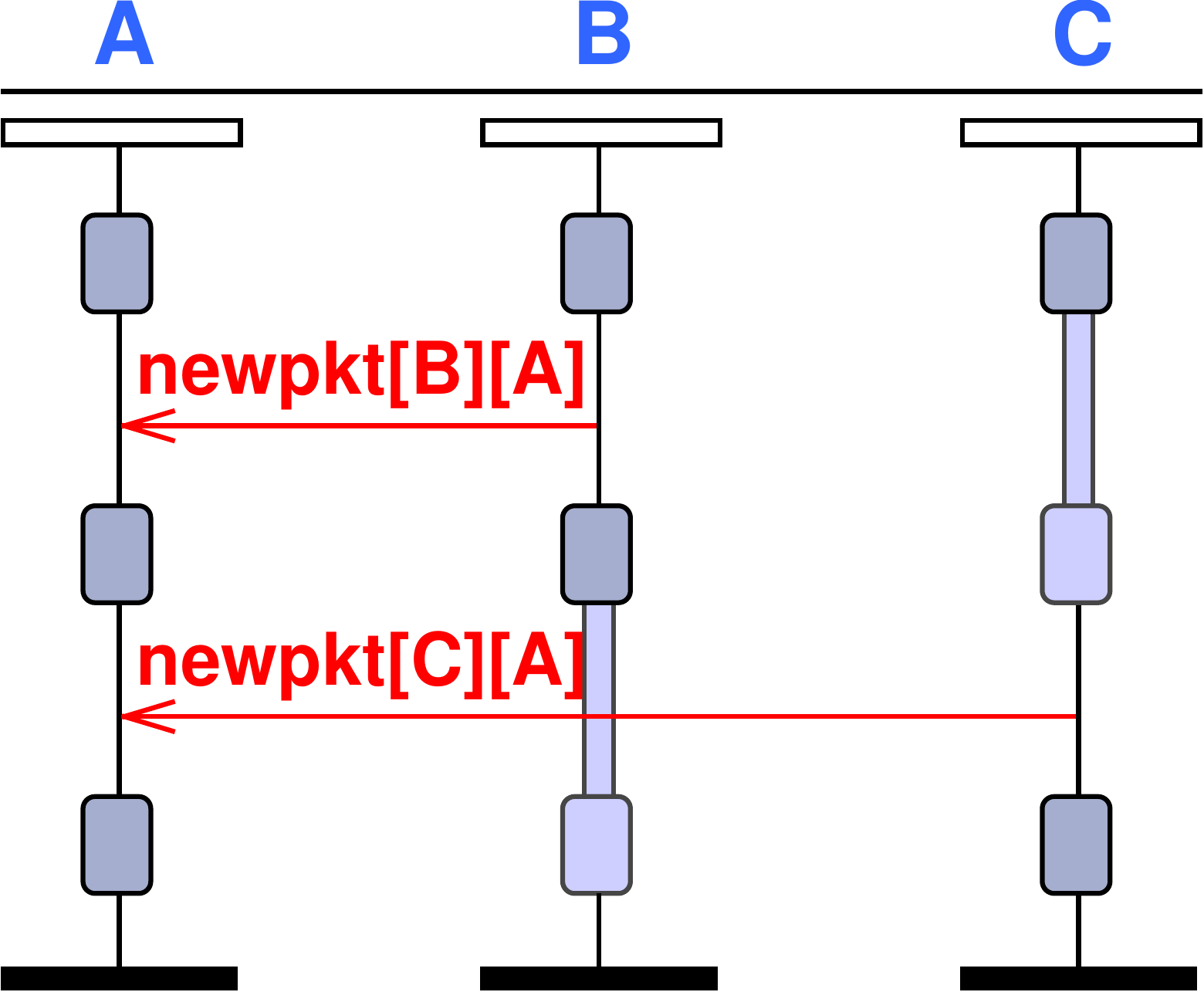}\hspace*{-1.5mm} 
}
\hfill\fbox{\hspace*{-1.5mm} 
\includegraphics[width=0.23\linewidth]{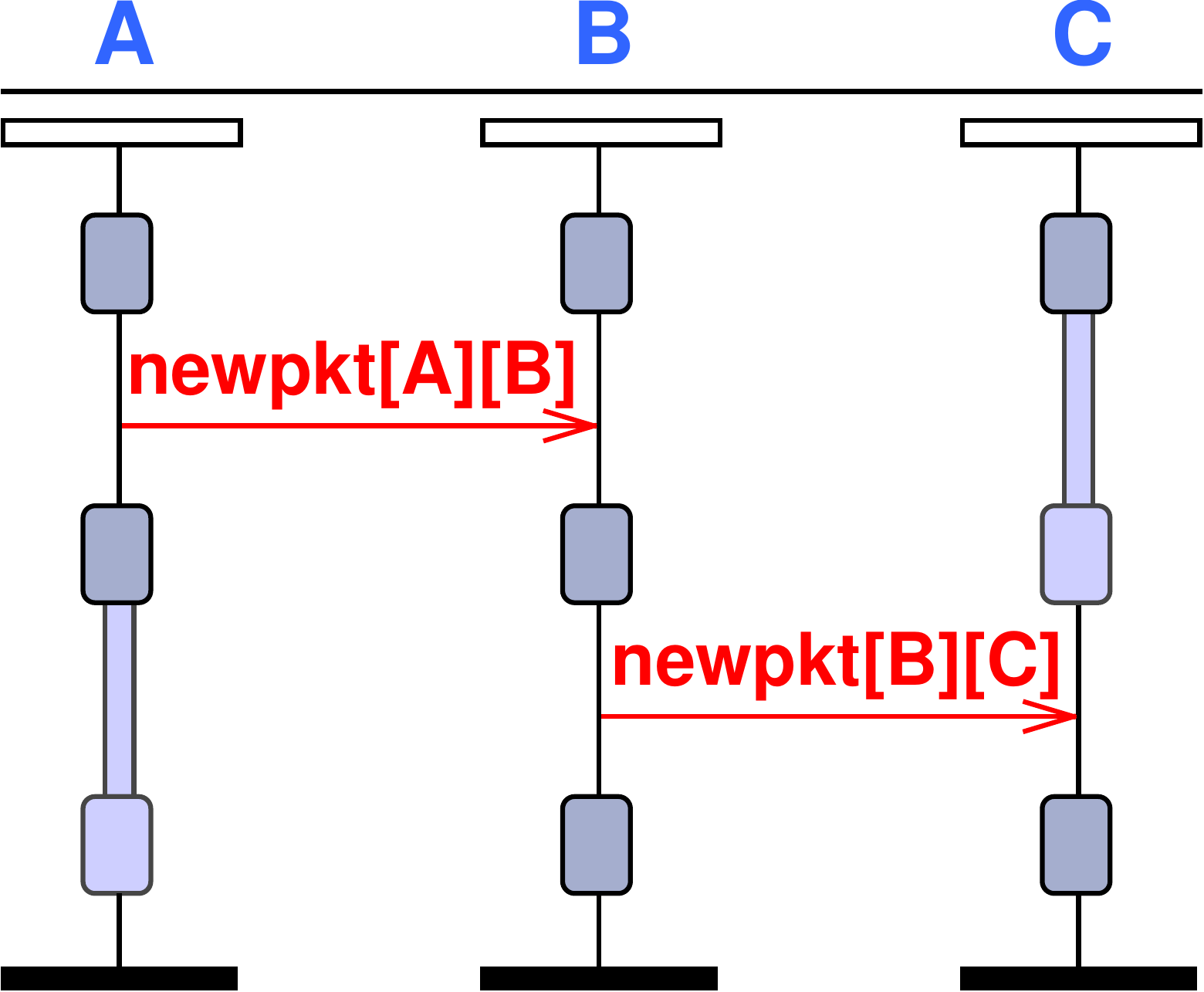}\hspace*{-1.5mm} 
}
\hfill\fbox{\hspace*{-1.5mm} 
\includegraphics[width=0.23\linewidth]{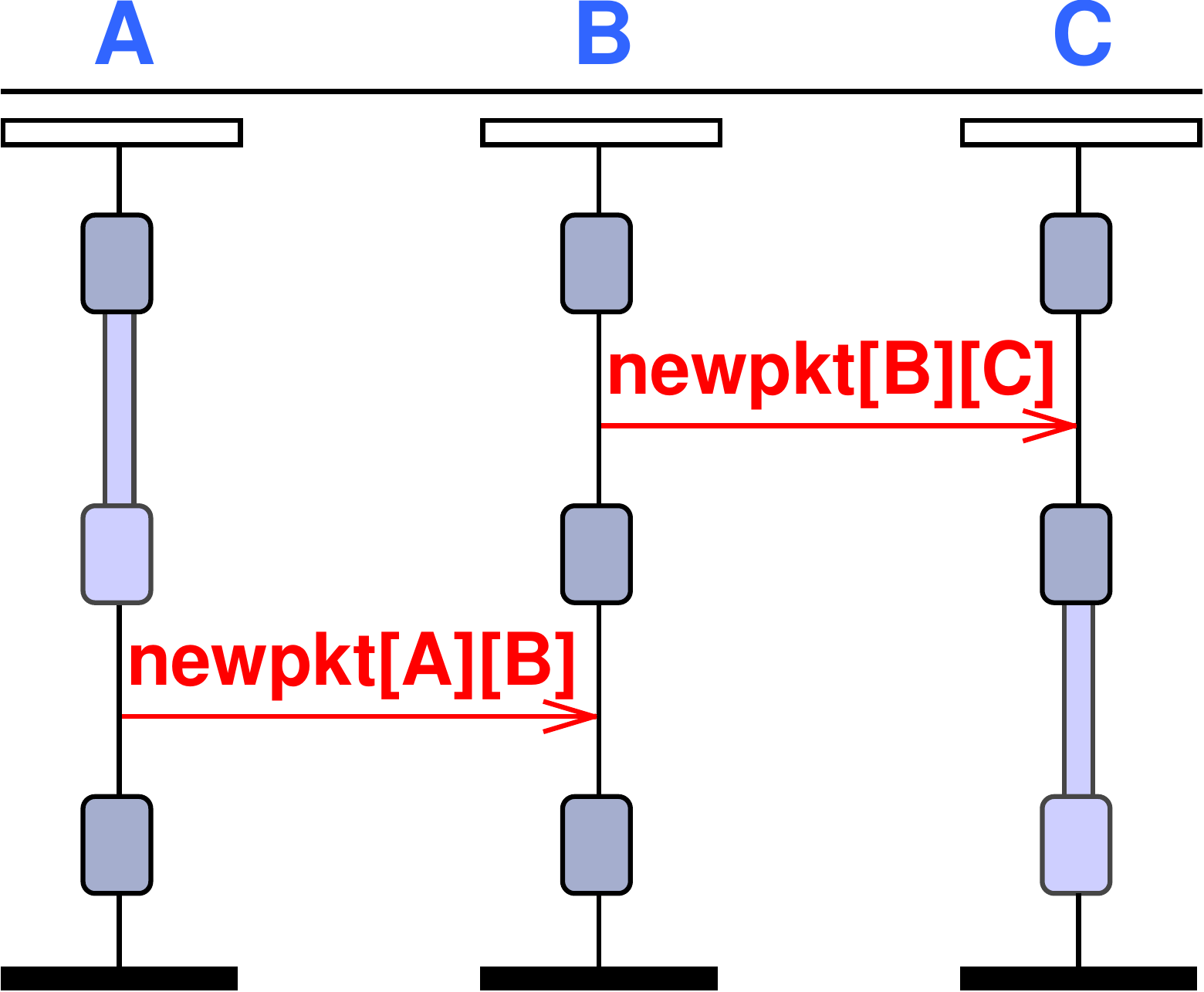}\hspace*{-1.5mm} 
}
\vspace*{-0.5em}
\caption{Sequence charts illustrating four scenarios for initiating two route requests.}\label{fig:scenarios}
\vspace{-1em}
\end{figure}
The new data packets may arrive as depicted in \Fig{scenarios}.
In the first scenario a packet from $A$ to $B$ is followed by a packet from $A$ to $C$;
in the second a packet from $B$ to $A$ by a packet from $C$ to $A$; 
in the third a packet from $A$ to $B$ by a packet from $B$ to $C$;
and in the final scenario a packet from $B$ to $C$ by a packet from $A$ to $B$.
The originator of the first new packet initiates a route discovery process first,  
the originator of the second non-deterministically after the first. The different scenarios are implemented by a simple automaton, \texttt{tester}.  Since the different topologies cover all possible permutations, these four scenarios cover all scenarios for injecting two new packets with either different originators or different destinations. 
\pagebreak[3]

Additional to $A$, $B$ and $C$, we add up to two nodes that may relay messages, but do not create data packets themselves. 
We consider only topologies in which nodes $A$, $B$ and $C$ are connected, either directly, or indirectly. This ensures that the route discovery is at least theoretically possible.
If it fails, then it won't be because the nodes are not connected, but due to failure of the protocol.

We consider three classes of topologies. 
The first class are static topologies. Given the constraints that node $A$, $B$ and $C$ are connected, and that there are at most 5 nodes, this gives $444$ topologies, after topologies that are identical up to symmetries are removed. 
The second class considers pairs of topologies from the first class, in which the second topology can be obtained by adding a new link. This models a dynamic topology in which a link is added. There are $1978$ such pairs. 
The third class considers the same pairs, but now moves from the second topology to the first. This models a link break. 
Note that after deletion, nodes $A$, $B$ and $C$ are still connected. 
In our \uppaal model a change of topology is modelled by another automaton. It may add or remove a link exactly once, non-deterministically, after the first route request arrives at the destination.

\subsection{Properties}

This paper considers three desirable properties of any routing protocol such as AODV. The first property is that once all routing messages have been processed a route from the originator to the destination has been found.
In \uppaal syntax this safety property can be expressed as:
\vspace{-2.7ex}

{\small \begin{align}\label{property1}
\begin{array}{r@{}l}
 \texttt{A[\,](}&\texttt{(tester.final \&\& emptybuffers()) imply}\\
&\texttt{(node(OIP).rt[DIP].nhop!=0))}
\end{array}
\end{align}}%

\noindent The CTL formula {\tt A[]$\phi$} is satisfied
if $\phi$ holds on all states along all paths. The
variable \texttt{node(OIP).rt} models the routing table of the originator node \texttt{OIP}, and the field
\texttt{node(OIP).rt[DIP]\!.nhop} represents the next hop for  destination
\texttt{DIP}.
All initiated requests will have been made, iff automaton
\texttt{tester} is in location \texttt{final}, the message buffers are
empty iff function \texttt{emptybuffers} returns \emph{true}, and
the          originator \texttt{OIP} has a route to node \texttt{DIP}
iff \texttt{node(OIP).rt[DIP].nhop!=0}.  

The second property is related, namely that once all messages are processed, then no sub-optimal route has been found. Here, sub-optimal means that the number of hops is greater than the shortest path. In case that the topology changes, we take the greater distance. In \uppaal this can be expressed as 
\vspace{-2.7ex}

{\small\begin{align}\label{property2}
\begin{array}{r@{}l}
 \texttt{A[\,](}&\texttt{(tester.final \&\& emptybuffers()) imply}\\
&\texttt{(node(OIP).rt[DIP].hops<=distance[OIP][DIP]))}
\end{array}
\end{align}}%

\noindent Here, the array \texttt{distance} encodes the distance matrix. Note, that this fails if the route at the end is sub-optimal. It does not fail if at the end, either an optimal, or no route has been found. If the first two properties are satisfied, it means that it is guaranteed that an optimal route will be found when all messages have been processed. Note that it is known that AODV does not guarantee that optimal routes will be found. Nevertheless, an implementation or modification of AODV can be said to perform better if this property fails for fewer topologies.

The third property is even stronger than the second, namely that no sub-optimal routes will be found at all. It does not hold if a better optimal route replaces a sub-optimal route that was found first. 
\vspace{-2.2ex}

{\small\begin{align}\label{property3}
 \texttt{A[\,](node(OIP).rt[DIP].hops<=distance[OIP][DIP])}
\end{align}}%

\noindent If the third property holds, then the second must hold as well. In the
experiments we will check all three properties for both
originator-destination pairs at once.

\subsection{Modifications}\label{ssec:modifications}
The basic \uppaal model is based on the process algebraic \awn model,
which reflects a common interpretation of the RFC with all ambiguities resolved.
It is known that AODV does not guarantee that optimal routes will be found, or even any routes at all~\cite{FGHMPT11,MK10}.\footnote{AODV proposes to repeat the route discovery process if the first discovery process fails. However, this solution does not solve the problems entirely (see~\cite{TR11}).} Our experiments quantify how many topologies are affected by these problems, and also what impact slight modifications of the protocol have. We will refer to the basic model as \emph{model 1}, and discuss three proposed variants of AODV.

\noindent\rule{0pt}{11pt}{\bf Forwarding all route replies.}
It is a known problem that nodes drop route reply messages under certain conditions.\footnote{This problem has already been raised on the MANET mailing list in Oct
2004 (\url{http://www.ietf.org/mail-archive/web/manet/current/msg05702.html}).} During our experiments we found this problem even in the smallest topology, a static linear topology with only three nodes, and only two links: node $A$ is connected to node $B$ and  $B$ to node $C$. Both node $B$ and $C$ initiate a route request to $A$. For this topology and scenario, \uppaal finds  a counterexample for Property~\eqref{property1}, i.e., it is possible that no route will been found when all messages have been processed.
\label{subsec:non-optimal-route}

\begin{figure}[t]
 \centering
\includegraphics[width=0.5\linewidth]{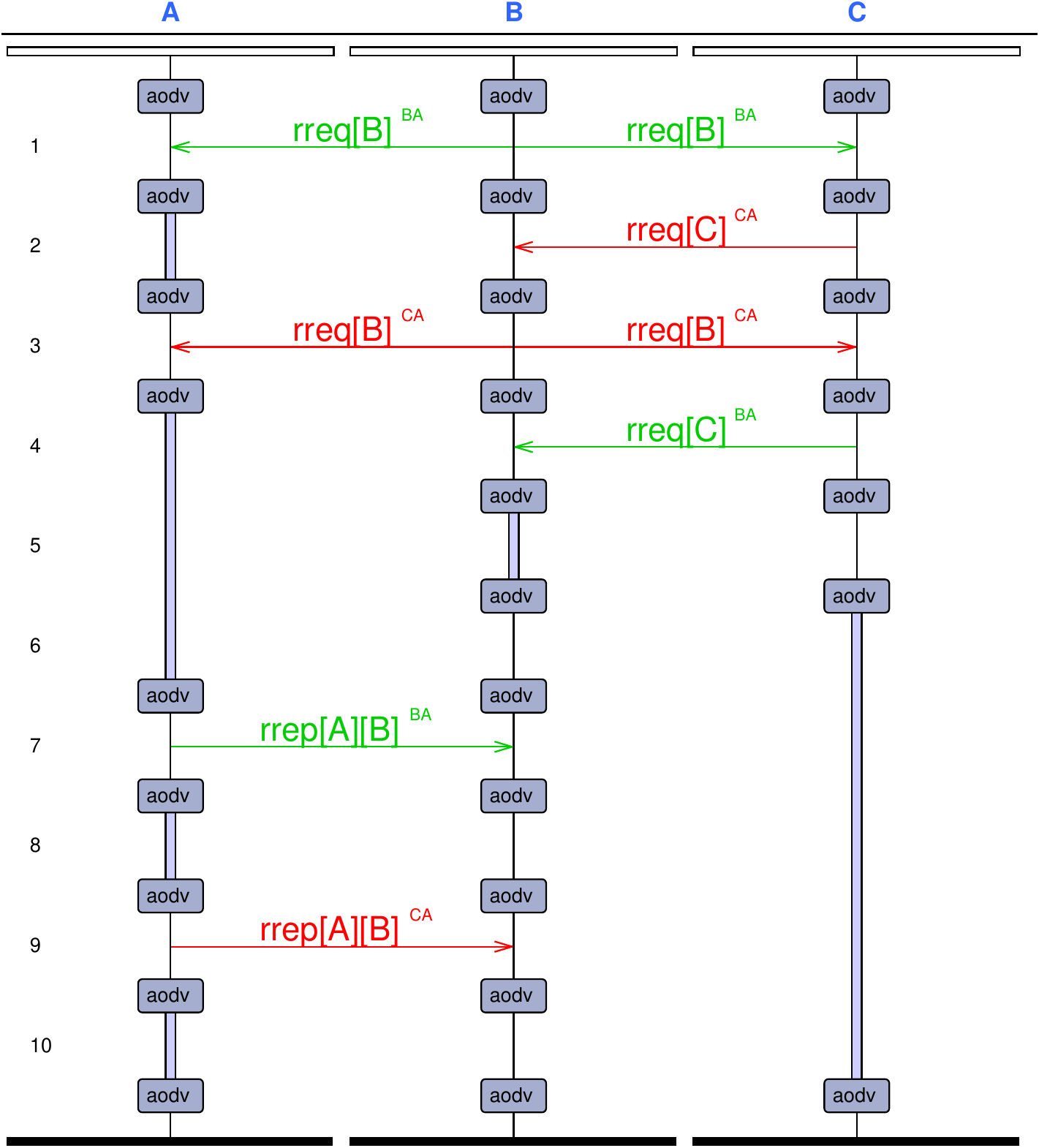}
 \caption{Message sequence chart illustrating failed route discovery. Wide vertical lines mean that local states do not change in this transition. The superscripts indicate the corresponding originator and destination of the route discovery process.}
 \label{fig:event}
 \vspace{-10pt}
\end{figure}
\newcommand{\ba}{\mbox{\it BA}}
\newcommand{\ca}{\mbox{\it CA}}
Fig.~\ref{fig:event} depicts a message sequence chart of the relevant part of the counterexample. Initially, both $B$ and $C$ initiate a route request for $A$. We refer to the first request as \ba-request, and to the second as \ca-request. First, node $B$ sends the \ba-request to $A$ and $C$ (Step 1 in Fig.~\ref{fig:event}), then node $C$ its \ca-request to $B$ (Step~2). Node $B$ forwards the \ca-request (Step 3), node $C$ the \ba-request (Step 4). Node $C$ will correctly ignore the \ca-request that it received from $B$, since it is the originator (Step 5). Similarly, $B$ will ignore the \ba-request (Step~6). Node $A$ will then reply to the \ba-request (Step 7), and node $B$ will update its routing table (Step 8) to include a route to $A$. Node $A$ will also reply to the \ca-request (Step 9), but $B$ will ignore this message (Step 10), since it does not contain new information for $B$. Node $A$'s reply to the \ca-request will not arrive at~$C$.

The discarding of the RREP message happens according to the RFC
specification of AODV \cite{rfc3561}. It states that an intermediate node
only forwards the RREP message if it is not the originator node
\emph{and} it uses the RREP to update its route entry to the destination. In this case, node $B$ is not the originator, but it also did not use the route reply to update its route. It already had an optimal route, as a result of the \ba-request. This type of problem can arise whenever one node has to relay multiple route requests for the same destination. 

A possible solution would be to forward every reply
received by a node. Our \emph{model 2} implements this change. Obviously, this increases the number of control messages generated during route discovery. 
However, this is compensated by the reduced need to repeat sending the route request
in case no route has been found, the solution proposed by AODV.\footnote{Moreover, a repeated route request 
need not be any more successful than the first.} In the experiment section we will see that this modification effectively addresses the problem.

\noindent\rule{0pt}{11pt}{\bf Replying to improving requests.}
Counterexamples found by \uppaal show that a 
source for sub-optimal routes is the property of AODV to only reply to the first route request. All subsequent requests with the same request ID (\texttt{rreqid}) will be ignored (Line 3 of Tables \ref{pro:rreq} and \ref{pro:xta}), 
even if the subsequent requests arrived via a shorter route.
\emph{Model 3} modifies the rule for the handling of route requests. It will not only reply to the first request, 
but also to a subsequent request (with the same request ID) with an improved hop count.

\noindent\rule{0pt}{11pt}{\bf Recovering from failed replies.}
Analysis of \uppaal's counterexamples show that a main reason
for failed route discovery is that a node marks a request as having
been replied to, even if the node detected the reply failed 
due to the link being broken in the time between the received request 
and the sent reply. The node will ignore other requests with the
same request ID that may arrive later. \emph{Model 4} introduces two
changes:  it does not mark a request as seen if the reply
fails, and it replies to
other requests in the same route discovery process. 

This change should be considered with care, since it changes the rules with respect to sequence numbers. These numbers are an essential part of AODV being loop free, and there is currently no guarantee that this change will not violate some essential invariants of the proof~\cite{TR11}. We included the results nevertheless, as they show that there is still significant potential to improve AODV.

\subsection{Experimental Results}

The experimental results tell for how many topologies \uppaal could show the absence of counterexamples, and  thus 
allow quantification of
the impact of improvements. However, the analysis uses a non-deterministic model, rather than a probabilistic model. For each topology it is reported whether a counterexample exists, but not how likely it is to occur. Neither can we assume that the topologies themselves are randomly distributed. Depending on the application only certain types of topologies might occur in practice. Nevertheless, it is fair to assume that a modification that leads to fewer topologies with counterexamples constitutes an improvement w.r.t. the considered property.

Table \ref{tab:results} presents the results of the experiments. Most relevant for all classes of topologies are Property~\eqref{property1}, a route is found, Property~\eqref{property2}, no sub-optimal~route is found in the end, and the combination of these, i.e., an optimal route is found.

The results demonstrate that the problem of ignoring route replies as described in \Fig{event} occurs even for about $50\%$ of all static topologies. \emph{Model 1} satisfies Property~\eqref{property2} only for half of all static topologies. The proposed modification solves this problem entirely for static topologies. The other modifications further improve the quality of the routes; in $99.1\%$ of static topologies Property~\eqref{property2} holds, i.e., the route was in the end always optimal. The slight drop in Property~\eqref{property3} is explained by the fact that in a few cases, where no route was found at all for \emph{model 1}, a sub-optimal route was found in the other models.

The results for static topologies are roughly repeated if we consider topologies in which a link is added. There were a few surprising instances though, in which adding a link was instrumental in finding a sub-optimal route. \pagebreak[3]

\setcounter{table}{2}
\begin{table}[t]
\begin{center}
\scriptsize
\begin{tabular}{|c|c|c|c|c|c|c|}
\cline{3-7}
\multicolumn{2}{c|}{\phantom{$(T^7)^7$}}&\, Property~\eqref{property1} &\, Property~\eqref{property2}\,&\, Property~\eqref{property3}\, &\, Property~\eqref{property1} \&~\eqref{property2}\, &\, all properties\, \\
\hline
\multirow{4}{*}{\begin{sideways}static\hspace*{3pt}\end{sideways}}
&\rule{0pt}{7pt}model 1& 52.7\% & 93.2\% & 50.7\%& 50.0\% & 13.5\% \\ 
&model 2& 100.0\% & 93.2\% & 47.5\%& 93.2\% & 47.5\% \\ 
&model 3& 100.0\% & 99.1\% & 47.5\%& 99.1\% & 47.5\% \\ 
&model 4& 100.0\% & 99.1\% & 47.5\%& 99.1\% & 47.5\% \\ 
\hline
\multicolumn{7}{c}{}\\[-0.8mm]
\cline{3-7}
\multicolumn{2}{c|}{\phantom{$(T^7)^7$}}&\, Property~\eqref{property1} &\, Property~\eqref{property2}\,&\, Property~\eqref{property3}\, &\, Property~\eqref{property1} \&~\eqref{property2}\, &\, all properties\, \\
\hline
\multirow{4}{*}{\begin{sideways}add link\hspace*{1.5pt}\end{sideways}}
&\rule{0pt}{7pt}model 1&57.5\% & 90.8\% & 49.1\%& 53.3\% & 18.1\% \\ 
&model 2 & 100.0\% & 90.6\% & 46.2\% & 90.6\% & 46.2\% \\ 
&model 3&100.0\% & 97.8\% & 46.2\%& 97.8\% & 46.2\% \\ 
&model 4&100.0\% & 96.3\% & 46.2\%& 96.3\% & 46.2\% \\ 
\hline 
\multicolumn{7}{c}{}\\[-0.8mm]
\cline{3-7}
\multicolumn{2}{c|}{\phantom{$(T^7)^7$}}&\, Property~\eqref{property1} &\, Property~\eqref{property2}\,&\, Property~\eqref{property3}\, &\, Property~\eqref{property1} \&~\eqref{property2}\, &\, all properties\, \\
\hline
\multirow{4}{*}{\,\begin{sideways}remove\hspace*{0.5em}\end{sideways} \begin{sideways}\hspace*{0.6em}link\end{sideways}\,}
&\rule{0pt}{7pt}model 1&26.7\% & 90.5\% & 59.7\%& 26.2\% &  6.0\% \\ 
&model 2 & 53.0\% & 89.4\% & 57.1\% & 51.2\% & 28.9\% \\ 
&model 3 & 53.0\% & 93.1\% & 57.1\% & 52.8\% & 28.9\% \\ 
&model 4 & 75.4\% & 94.0\% & 54.0\% & 73.8\% & 41.0\% \\ 
\hline \multicolumn{7}{c}{}
\end{tabular}
\caption{Model checking result for the four models and three classes of topologies. It gives the percentage of topologies for which there exists no counterexample.}\label{tab:results}
\end{center}
\vspace{-3.2em}
\end{table}

The results are, as expected, not quite as positive if a link gets removed.
For the baseline model it is only guaranteed for one quarter of all topologies that a route will be found. Relaying all route replies, and not marking requests if the reply fails, improves this result. For three quarters of all topologies in which a link was removed it was shown that an optimal route will be found. 

The main reason of the failures that remain is that a route reply
might get lost because of some intermediate link break on the path
back to the destination.  A possible solution to this problem
could be to maintain a set of back-up routes, or to implement  different error responses.  However, this requires a significant change and fundamentally changes the characteristics of AODV.
  
For the experiments we used an Intel Core2 CPU  2.13GHz processor with 2GB internal memory, running Ubuntu 11.04. We used \uppaal 4.0.13. Of each of the models described in this section, we checked $17600$ instances, altogether $70400$ instances. As indication of the state space and runtimes, we checked an invariant on all instances of \emph{model 4} for a topology in which a link is removed. These instances have larger state spaces than others, since these scenario have also to trigger the transitions for error handling. The models have an average of $9400$ states, the largest model has $475000$ states, and the median is $2700$. Exploring these state spaces took on average $1.73$ seconds user time, at most $81$ seconds, and the median was $0.57$. These run times show that an automated, systematic and rigorous analysis of reasonable rich routing protocols is feasible.

\section{Related Work}
\label{sec:related}

Other researchers have used formal specification and analysis techniques to investigate the correctness and performance of AODV; we survey the sample  related to model checking.

Bhargavan et al.\ \cite{BOG02} were amongst the first to use model checking on a 
draft of AODV, demonstrating the feasibility and value of automated verification of routing protocols. Their investigations using the SPIN model checker revealed that in some circumstances routes containing loops can be created. The proposed variation which guarantees loop freedom were not included in the \mbox{current standard.}

Musuvathi et al.\ \cite{MPCED02}  introduced the CMC  model checker
primarily to search for coding errors in implementations of protocols
written in C\@. They use AODV as an example and, as well as discovering
a number of errors, they also found a problem with the specification
itself which has since been corrected.

Chiyangwa and Kwiatkowska \cite{CK05} use the timing features of
UPPAAL to study the relationship between the timing parameters and the performance of route discovery. They established  a
dependence between the lifetime of a route and the size of the network, although their study only considered
the initiation of a single route discovery process,
and a static linear topology. In~\cite{FGHMPT11}, we confirmed some of the problems they discovered, and 
show their independence of time.

Other researchers have used model checking to analyse other routing protocols. Wibling et al.\ \cite{WPP04} for example used  SPIN and UPPAAL to verify aspects of the LUNAR protocol, which is also used in ad hoc routing for wireless networks. In particular the timing feature of UPPAAL was used to check upper and lower bounds on route finding and packet delivery times. The scenarios considered included a limited number of topology changes where problems were suspected.

De Renesse and Aghvami \cite{RA04} used SPIN  to study the WARP protocol.  To reduce the overhead on model checking, various simplifications were imposed on a five-node network, including a single source and destination and limitations on the degree that the network can change.

Fehnker et al. have used the model checker \uppaal to analyse a TDMA time synchronisation protocol \cite{FHM07}. Similarly to our approach they considered all topologies in a certain class, but did not cover dynamic topologies. 

Our approach is in line with these related works.
However, it is unique in the sense that our UPPAAL model complements
our process-algebraic specification of AODV. As mentioned before,  these two approaches to formal protocol modelling, specification and evaluation, if used together,
can provide a powerful tool for the development and rigorous evaluation of new protocols and variations, and improvements of existing ones.
Currently, our UPPAAL model is derived by hand directly from the \awn specification, but an automatic translation from \awn in the style of
Musuvathi et al. \cite{MPCED02} is possible, and remains as future work.

\section{Conclusions and Outlook}\label{sec:conclude}

The aim of this ongoing work is to complement by model checking
a process algebraic description of WMN routing protocols in general, and AODV in particular.
The used description of AODV described in \cite{TR11} is amongst the
first detailed formal models. Having the ability of automatically deriving 
an \uppaal model from an \awn specification and thus model checking
formal specifications allows the confirmation and detailed
diagnostics of suspected errors.
The availability of an executable model becomes especially useful
in the evaluation of proposed improvements to AODV, as we have shown.

We have sketched  possible modifications of AODV, which have been evaluated by formal and rigorous analysis by means of
model checking. An analysis of these modifications by means of process algebra is part of future work.
We have set up an environment where we can
test a whole bunch of different topologies in a systematic manner. This will allow us to do a fast comparison between standard AODV and proposed variations in contexts  known to be problematic.

\bibliographystyle{splncs03}
\bibliography{aodv}
\end{document}